\documentclass[12pt]{iopart}
\begin{document}

\newcommand{\E}{{\cal E}}
\newcommand{\F}{{\cal F}}
\newcommand{\A}{{\cal A}}
\newcommand{\alr}{\alpha r}
\newcommand{\asr}{\alpha^{2}r^{2}}
\newcommand{\dspst}{\displaystyle}

\title{Spherical and planar three-dimensional anti-de Sitter
 black holes}
\author{Vilson T. Zanchin\footnote{e-mail: zanchin@ccne.ufsm.br}
 and Alex S. Miranda\footnote{e-mail: amiranda@mail.ufsm.br}}
\address{Departamento de F\'{\i}sica, Universidade Federal
de Santa Maria,\\ 97119-900 Santa Maria, RS, Brazil}


\begin{abstract}
\noindent

 The technique of dimensional reduction was used in a recent
paper \cite{zkl} where a three-dimensional (3D) Einstein-Maxwell-Dilaton
theory was built from the usual four-dimensional (4D) Einstein-Maxwell-Hilbert
action for general relativity. Starting from a class of 4D toroidal black holes
 in asymptotically anti-de
Sitter (AdS) spacetimes several 3D black holes were obtained and
studied in such a context. In the present work we
choose a particular case of the 3D action which presents Maxwell
field, dilaton field and an extra scalar field, besides gravity
field and a negative cosmological constant, and obtain new 3D 
static black hole solutions whose horizons may have spherical
or planar topology.  We show that there is a
3D static spherically symmetric solution analogous to the 4D
Reissner-Nordstr\"om-AdS black hole, and obtain other new 3D
black holes with planar topology. 
From the static spherical solutions,
new rotating 3D black holes are also obtained and
analyzed in some detail.

\end{abstract}

\maketitle

\section{Introduction}
\label{intr}

The early works in three-dimensional (3D) general relativity showed
that it provides a test-bed  to 4D and higher-D theories \cite{achucarro,
witten,deser1,deser2,giddings}. Moreover, the work on 3D gravity theories
 has seen a great impulse after the discovery
that 3D general relativity possesses a black hole solution, the
BTZ black hole \cite{btz,bhtz}. This black hole is a solution of
the Einstein-Hilbert action including a negative cosmological
constant term $\Lambda$.  One can also show that the BTZ black
hole can be constructed by identifying certain points of the 3D
anti-de Sitter (AdS) spacetime \cite{bhtz,carlip}.

The BTZ solution arouse the interest on the subject of 3D black holes
and a  whole set of new solutions in 3D followed from a number of
different dilaton-gauge vector theories coupled to
gravity. For instance, upon reducing 4D Einstein-Maxwell theory with
$\Lambda$ and with one spatial Killing vector it was shown in
\cite{lemos1,lz1} that it gives rise to a 3D Brans-Dicke-Maxwell
theory with its own black hole, which when reinterpreted back in 4D
is a black hole with a toroidal horizon. One can then naturally extend
the whole set to Brans-Dicke theories \cite{sa,dias1}. Other solutions
with different couplings have also been found \cite{chan,martinez} (see
\cite{dias1} for a more complete list).

A renewal of interest in 3D black hole solutions in asymptotically AdS
spacetimes came after the
AdS-conformal field theory conjecture (AdS/CFT) \cite{maldacena}.
ADS/CFT tells us how to compute at strong coupling in quantum 
field theories using the near-horizon geometry of certain black branes.  
The 3D  black hole solutions we investigate in the present paper
can be important in the verification of the conjecture. It is known that
the BTZ black hole play an important role in this context,
 since many higher-D extreme black
holes of string theory have a near-horizon geometry containing the BTZ
black hole \cite{maldacenastrominger}. Other 3D black holes may also
be important in connection to  higher-D string theory such that they
can be interpreted as the near horizon structures of a brane
rotating in extra dimensions \cite{duff}. The solutions we study here 
are generalizations of the 3D black holes found in
reference \cite{zkl} including rotation, and have the BTZ black hole
as a particular case.

The motivation
to study lower dimensional black holes is also provided by the evidence
that some physical properties of black holes such as Hawking radiation and
evaporation can be more easily understood in the context of two and three
dimensions \cite{harvey,cadoni}. As we shall see, the 3D black holes
we study below have geometric properties very close to some 4D solutions
such as the Reissner-Nordstr\"om black hole. The study of these 3D black holes
may then help us to better understand the ordinary black hole physics.

On the other hand, dilaton gravity theories are of particular interest
since they emerge
as low-energy limit of string theories. They are commonly studied in
lower dimensions because the lower-dimensional Lagrangian yields
field equations whose solutions are easily found and analyzed. 

The usual starting point to find  solution in a given dimension, $D$ say,
is  the action of gravity theory with the generic
dilaton and gauge vector couplings in that dimension. One then
derives the corresponding equations of motion, tries an ansatz for the
solution and then finally from the differential equations finds
the black hole solutions compatible with the ansatz.

From the known solutions in that dimension one can find new solutions
in the same dimension if the theory possesses dualities  symmetries that
convert one solution into another in a nontrivial way
\cite{horowitzreview92}.

New solutions  in lower dimensions  can be obtained  through
dimensional reduction, where one can reduce a theory by several
dimensions. Among the  procedures to perform a
dimensional reduction, the  classical Kaluza-Klein
reduction \cite{duff2}, and the Lagrangian dimensional reduction
\cite{cremmer} are commonly used, and are both equivalent
when  reducing through one symmetric compact direction
(see e. g. \cite{horowitzreview92,duff2,cremmer,breiten,KSS,abicaks}).

In reference \cite{zkl} the  classical Kaluza-Klein procedure was used
to find new black hole solutions in 3D.   A 3D Lagrangian including
gravity, cosmological constant, dilaton field, Maxwell gauge field,
Kaluza-Klein gauge field, and an extra scalar field was obtained
from the 4D Einstein-Maxwell gravity theory.
Such a  3D Lagrangian is a Brans-Dicke-Dilaton-Maxwell gravity
theory with the dilaton playing a special role. In particular, the
dimensional
reduction through the Killing azimuthal direction
$\partial/\partial\varphi$, of rotating charged black holes with
toroidal topology was considered.
 The resulting 3D black hole displays an isotropic (i.e., circularly
symmetric) horizon, with two gauge
charges and a number of other interesting properties.

On could also expect to obtain a second class of three-dimensional
spherical and even anisotropic (non-spherical) black
hole solutions through the dimensional
reduction of 4D spacetimes such as the Kerr-Newman-AdS
family of black holes. The reduction can be performed through the Killing
azimuthal direction $\partial/\partial\varphi$ and, because of the special
topology of the Kerr-Newman-AdS black hole,  an analytical continuation
through the polar angle $\theta$ has to be made.
However, this procedure conduces to
ill-defined solutions that do not have  a natural interpretation from a
three-dimensional point of view.  These kind of problems can be avoided
by following different procedures as done for instance in \cite{frolov},
where a 3D distorted black hole was obtained through dimensional reduction
of the Schwarzschild black hole.

Other way of avoiding the above mentioned problems of topology is
starting from the 3D dilaton-gravity action, obtained through the
dimensional reduction
from 4D Einstein-gravity, and working entirely in 3D dimensions.
This is the strategy followed in the present work, where a
particular case of the 3D dilaton-action written in \cite{zkl}
 is used to build different classes of solutions.

In the next section it is presented a summary of
the model, the action and the field equations are written,
 and the definitions of global charges are reviewed.

Assuming that the spacetime admits one timelike Killing vector
and one spacelike Killing vector, the metric is then static 
and has a spatial symmetry that can be spherical or planar
depending on the topology of the spacetime. We
make an ansatz for the other fields that are allowed to
be non-spherical. This is done in section \ref{sectsolutions},
by using Schwarzschild-like  coordinates $(t, r, \theta)$,
 and assuming that the metric depends only on the radial coordinate $r$,
while the other fields depend also on the coordinate $\theta$.
It is then showed that there are solutions to the field equations
representing spacetimes endowed with spherical black holes and
with non-spherical dilaton field.

Five different classes of solutions presenting black holes  are
 reported in the following sections.   We show that the solution
presented in section \ref{sectspherical} is in fact a particular
case of the 3D black hole studied in \cite{zkl}. The other 
cases are new 3D black hole solutions.  A planar black hole is obtained
 in section \ref{sectplanar}
by assuming the coordinate $\theta$ to be non-periodic. 
The spherically (circularly) symmetric  spacetime 
analyzed in section \ref{sectrnads} is similar to the 4D
Reissner-Nordstr\"om-AdS black hole.  
Considering a different topology of the subspace with constant values of
$r$ and $t$, the same solution analyzed in
section \ref{sectrnads} may be interpreted as a planar black hole. This is  
briefly discussed in section \ref{sectplanar1}. 
Other planar black hole
solution is presented in section  \ref{secthyper}.
The rotating versions of the new 3D spherical
 black holes are considered  in section \ref{sectrotating}.
In section \ref{sectconclusions} we conclude.

\section{3D Dilaton-Maxwell gravity}
\label{lagrangian}

\subsection{The 3D Lagrangian and field equations}

The Kaluza-Klein dimensional reduction of 4D
black holes was used in  \cite{zkl} to build new 3D spacetimes
with black holes.
In this section we summarize the results obtained there
for static black holes and review the interpretation issue
of the reduced solutions as 3D black holes.

It is assumed that the 4D manifold ${\cal M}_4$ can be decomposed as
${\cal M}_4={\cal M}\times S^1$, with
$\cal M$, $S^1$, being the 3D manifold and the circle, respectively.
For the sake of simplicity
we particularize the discussion to the case of static 4D spacetimes.

The four-dimensional Lagrangian $\hat{\cal L}$ is assumed to be the usual
Einstein-Hilbert-Maxwell Lagrangian with cosmological term $\hat
\Lambda$, and electromagnetic field $\hat {\bf F} = {\rm d} \hat {\bf
A}$ (where $\hat {\bf A}$ is the gauge field), given by  (we use geometric
units where $G=1$, $c=1$).
\begin{equation} \hat S = \int{dx^4\, \hat{\cal L}}=
\frac{1}{16\pi}\int{d^{4}x\sqrt{-\hat g} \left(\hat R-2 \hat \Lambda -\hat
{\mathbf F}^2\right)}\, .                         \label{lagr4d}
\end{equation}
The convention adopted here is that quantities wearing hats are defined in
4D and quantities without hats belong to 3D manifolds.

 In order to proceed with
the reduction of the Lagrangian $\hat{\cal L}$
to 3D, consider  a 4D spacetime
metric admitting one spacelike Killing vector, $\dspst{\partial_\varphi}$,
where $\varphi$ is assumed to be a compact direction. In
such a case the 4D static metric may be decomposed in the form
\begin{equation*}
d{\hat s}^2= \Phi^{2\beta_0}ds^2
+\Phi^{2\beta_1} d\varphi^2 \, ,
\end{equation*}
where $ds^2$ is the 3D metric, $\Phi$
and all the other metric coefficients are
functions independent of $\varphi$, and $\beta_0$ and $\beta_1$ are real
numbers. 

We are free to choose $\beta_0$ and $\beta_1$, i.e., we are free
to choose the frame in which to work, with the different frames being
related by conformal transformations.  As discussed in detail in reference
\cite{zkl}, the most interesting frame is the  one that fixes
$\beta_0 =0$ with $\beta_1$ free. In this paper we choose $\beta_1 =1$.
The dilaton field $\Phi^2$ is assumed to be nonzero, except 
at isolated points where it may be zero.

It is well known that different frames, related by conformal
transformations, are physically inequivalent, e.g., one frame can give
spacetime singularities where the other does not (see, e.g.,
\cite{chan-mann,marolf,fiziev}). 
  Therefore, once a choice of parameters 
$\beta_0$ and $\beta_1$ has been made, then a particular frame has been chosen,
and the metric in this frame is interpreted as describing the physical spacetime.
Other choices of parameters lead to solutions whose spacetime may
have different physical and geometric properties.

To dimensionally reduce the electromagnetic gauge field we do
\begin{equation*}
\hat {\bf A} = {\bf A} + \Psi {\rm\bf d}\varphi\,,
\end{equation*}
where for compact $\varphi$ the gauge group is $U(1)$.
 In the last equation ${\bf A}$ is a
$1$-form while $\Psi$ is a $0$-form.
From this we can define the 3D Maxwell field
\begin{equation*}
\bf F = {\rm {\bf d}}\bf A\,.
\end{equation*}

The Kaluza-Klein dimensional reduction procedure then gives
\begin{equation}
S = \frac{L_3}{16\pi }\int{{\cal L}\, d^3 x }=
\frac{L_3}{16\pi }\int {d^{3}x\sqrt{- g \Phi^2\,}
\left[R+6\alpha^2-{\mathbf F}^2 -{2}\Phi^{-2} ({\bf \nabla} \Psi)^2\right]}\, ,
\label{lagr3d1} \end{equation}
where  $L_3$ is the result of integration
along the $\varphi$ direction. The 3D cosmological constant is defined
as $\Lambda=\hat \Lambda$, and we have written $\alpha^2 \equiv- \Lambda/3$.

 The equations of motion which follow from the action (\ref{lagr3d1}) for the
graviton ${\bf g}$, the gauge field ${\bf A}$, the dilaton $\Phi$,
and the scalar $\Psi$ are, respectively,
\begin{eqnarray}
&  & G_{ij}= 3\alpha^2g_{ij} + \Phi^{-1}
\left(\nabla_i \nabla_j\Phi - g_{ij}\nabla^2\Phi\right)
 +2\left( F_{ik}{ F_j}^{k} - {1\over 4}g_{ij}{\mathbf F}^2 \right) \nonumber \\
& &\hskip 1.3cm +{2}\Phi^{-2} \left(
\nabla_i\Psi\nabla_j\Psi  -{1\over 2}g_{ij}(\nabla\Psi)^2\right)
 \, , \label{einsteineq}\\
& & \nabla_j \left(\Phi \,F^{ij} \right) =0 \label{gaugefieldeom} \, ,\\
 & &\nabla_i\left(\nabla^i\Phi\right)  =
+3\alpha^2\,\Phi +  {1\over 2} \Phi\,F^2
 - \Phi^{-1}\left(\nabla\Psi\right)^2 \, , \label{phieom} \\
& & \nabla_i\left[\Phi^{-1}\nabla^i\Psi\right] = 0\, ,\label{psieom}
\end{eqnarray}
where $G_{ij}$ is the Einstein tensor.

\subsection{The global charges}
\label{sectcharges}
For the sake of convenience, we write here the definitions of
gravitational mass, angular momentum,
and charges in the 3D spacetime arise by using
the formalism of Brown and York \cite{by1}  modified to include a
dilaton and other fields \cite{brown-mann,creighton-mann, booth-mann}.
The conventions adopted  are the same as in \cite{zkl}.\\

\noindent{\it (i) Conventions}:
We assume that the 3D spacetime  $\cal M$ is topologically the
product of a spacelike surface $D_2$ and a real line time interval $I$,
${\cal M} =D_2\times I$.  $D_2$ has the topology of a disk. Its boundary
$\partial D_2$ has the topology of a circle and is denoted by ${\cal S}_1$.
The boundary of $\cal M$, $\partial \cal M$, consists of
two  spacelike  surfaces $t=t_{1}$ and $t=t_{2}$, and a timelike
surface ${\cal S}_1\times I$ joining them. Let $t^i$ be a timelike
unit  vector ($t_it^i=-1$)
normal to a spacelike surface $D_2$ (that foliates $\cal M$), and $n^i$
be the outward unit vector normal to the boundary $\partial\cal M$
($ n_in^i=1$). Let us denote the
spacetime  metric on $\cal M$ by  $g_{ij}$ ($i,j=0,1,2$).
 Hence $h_{ij}= g_{ij} + t_it_j$ is the induced metric on $D_2$
and $\sigma_{ij}= g_{ij}+t_it_j-n_i n_j$ is the  induced metric on
${\cal S}_1$. The induced metric on the spacetime
boundary $\partial \cal  M$ is $\gamma_{ij}= g_{ij}-n_i n_j =
\sigma_{ij}-t_i t_j$. We also assume that the spacetime admits the
two Killing vectors needed  in order to define mass and angular
momentum: a timelike Killing vector
$\eta_{t}^{i}=(\frac{\partial}{\partial t})^{i}$ and a spacelike
(axial) Killing vector
$\eta_{\theta}^{i}=(\partial_\theta)^{i}$. \\

\noindent{\it (ii) Mass}:
By adapting the Brown and York procedure to take into
account the dilaton field the following
definition of mass $M$ on a 3D spacetime admitting a timelike Killing
vector $\eta_t$ is obtained
\begin{equation} M = -{L_3}\int_{{\cal S}_1}
\delta\left({\epsilon^\Phi}\right)t_{i} \eta_t^{i}\,d{\cal S}\, ,
\label{massdef} \end{equation} where  $d{\cal S} =\sqrt{\sigma}\,d\theta$
with $\theta$ being a coordinate on ${\cal S}_1$, and $\sigma$ being the
determinant of the induced metric on ${\cal S}_1$.
 $\epsilon^\Phi$ is
energy surface density on ${\cal S}_1$. Recall that $\epsilon^\Phi
=\frac{\kappa^\Phi}{8\pi}\,$, where $\kappa^\Phi$
 is the trace of the extrinsic curvature of ${\cal S}_1$ as
embedded on $D_2$, modified by the presence of the dilaton (see  below).\\

\noindent{\it (iii) Angular momentum}:
Similarly to the  mass, the definition of  angular momentum $J$
for a 3D spacetime admitting a spacelike Killing vector
$\partial_\theta=\eta_\theta$ can also be modified to include the dilaton.
For the cases we are going to consider here, a sufficiently general
definition is
 \begin{equation}
J= {L_3}\int_{{\cal S}_1} \delta\left({j^{\Phi}}_{i}
\right)\eta_\theta^{i} \,d{\cal S}\, ,
\label{momentumdef}
\end{equation}
where ${j^{\Phi}}_{i}$ is the momentum
surface density on ${\cal S}_1$, modified by the presence of the dilaton.
  The angular momentum density may be written as $
\displaystyle{
j^\Phi_i = -2{\sigma_{ij}} n_l {P_{\Phi}}^{jl}/\sqrt{h},}$
where $P_\Phi^{jl}$ is the canonical gravitational momentum, modified by the
presence of the dilaton \cite{creighton-mann}. The other quantities
were defined above, and are viewed as tensor quantities
in the two-space $D_2$.

To define ${\kappa^\Phi}$ and $j^\Phi_i$ explicitly we consider the
 case when the  3D metric  can be split as
\begin{eqnarray}
\fl ds^2& =& -N^2dt^2+ h_{mn}\left(dx^m+ N^m
dt\right)\left(dx^n+N^n dt\right) \,\nonumber\\
\fl &=&- \left(N^2-R^2(N^\theta)^2\right) dt^2 +  2R^2 N^\theta d\theta\, dt
+  f^{-2}dr^2 +  R^2\left(d\theta+V\,dr\right)^2\,, \label{splitmetric}
\end{eqnarray}
where $m,n=1,2$, $x^1 =r$, $x^2=\theta$, and $\theta$ parameterizes ${\cal
S}_1$. Functions $N$, $N^\theta$, $f$, $R$ and $V$ depend on
coordinate $r$ only.
$\kappa^{\Phi}$ may then be written  as  \begin{equation} {\kappa^{\Phi}}=
-{1\over 2}{f} \left[{2 } {\partial\Phi\over\partial r} + \Phi\left(
{2\over R} {\partial R\over\partial r}
 -\nabla_\theta V \right)\right]\, ,
\label{kappadef} \end{equation}
where $\nabla_\theta V$ is the covariant derivative on ${\cal S}_1$, which
in the present case is identically zero.

Now, the angular momentum density is given by
\begin{equation}
j_i^\Phi\eta_\theta^i = {1\over 8\pi}{\partial N^\theta\over\partial r}
{\,R^2\over N}\sqrt{f\Phi^2\left(1+R^2V^2f\right)}\, .
\label{jaydef} \end{equation}

\vskip .6cm
\noindent {\it (iv)  Electric charge  of the gauge field  $A_i$:}
The gauge charge can be
obtained by the Gauss law, adapted to non-asymptotically
flat stationary spacetimes \cite{by1}, and to the presence of the
dilaton (see also \cite{lemos1,lz1})
\begin{equation}
{Q_e} ={L_3\over 4\pi}\int_{{\cal S}_1}{\delta \left(E_i\right)n^{i}}
d{\cal S} \,, \label{echarge1}
\end{equation}
where $E_i\equiv \Phi\, F_{ij} t^{j}$.

The magnetic charge is not defined here since the 3D spacetimes we
consider below do not contain magnetic Dirac-like monopoles.\\

\noindent{\it (v) Dilatonic charges}:
 Stationary asymptotically AdS 3D black holes may also be
 char\-ac\-ter\-ized by the dilaton charge. In fact, two different
definitions can be formulated \cite{gibbonsperry,ghs}.
 However,  dilaton charges are of little importance in our analysis
because, for the particular dilatonic black holes considered here,
they are identically zero.\\

\noindent{\it (vi) Charges of the scalar field $\Psi$}:
 In this case,
the equation of motion for $\Psi$ yields two  conserved currents.
It is then possible to define the  corresponding
conserved  charges (of electric and magnetic type, respectively) as
\begin{equation}
Q_{\Psi}=
{L_3\over 4\pi}\int_{{\cal S}_1}
\delta\left({1\over \Phi}\nabla^i\Psi\right)\, n_{i}\,d{\cal S} , \label{qpsi}
\end{equation}
\begin{equation}{\tilde Q}_{\Psi} = {L_3\over 4\pi} \int_{{\cal S}_1}   
    {\epsilon_{ijk}\delta\left(\nabla^k\Psi \right)\, t^in^j}d{\cal S} \, .
\label{qpsidual}
\end{equation}
It is straightforward to verify that both of the above definitions
represent conserved charges. {F}rom equation (\ref{psieom}) it is
possible to identify the quantity $J_{\Psi}^i =
{1\over\Phi}\nabla^i\Psi$ as a conserved current, $\nabla_i
J_\Psi^i =0$. The  corresponding conserved  charge (analog to the
electric charge) being (\ref{qpsi}). The second charge
(\ref{qpsidual}) is also conserved for
 it follows from the conserved current
$\nabla^j\left(\epsilon_{ijk }
\nabla^k\Psi \right)$.

\section{3D static black hole solutions}
\label{sectsolutions}

The dimensional reduction of 4D Kerr-Newman-AdS black-holes suggests we
can find new 3D spherical (and non-spherical) black hole solutions. Assuming
the spacetime is static, i. e., there is a timelike Killing vector,
we can choose Schwarzschild-like coordinates ($t, r, \theta$) 
in which the metric can be written in the diagonal
form as showed in the Appendix, whose coefficients depend
on coordinates $r$ and $\theta$ only.

If  we restrict the metric coefficients to depend only upon
one of the radial coordinate $r$, i.e., the 3D spacetime admits also
a spacelike Killing vector, the resulting metric 
can be  written as
\begin{equation}
ds^2 = -F(r)dt^2 + {1\over F(r)}dr^2 +r^2d\theta^2 \label{metric1} \, ,
\end{equation}
where $F(r)$ is an arbitrary function.

Coordinates $t$ and $r$ are defined in the ranges: $-\infty<t<+\infty$, 
and $0\leq r<+\infty$. Therefore,
the above metric contains two different topologies of interest
in the present work, depending on the topology of 
$C_\theta$ the hyper-surfaces (lines) with constant values of $t $ and $r$. 
If $\theta$ is a compact 
direction ($\theta$ is an angular coordinate), $C_\theta$ is the circle $S^1$,
and (\ref{metric1}) represents a spherically symmetric spacetime.
On the other hand, if $\theta$ is a non-compact 
direction ($-\infty< \theta<+\infty$), $C_\theta$ is the real line ${\bf R}$
and the spacetime is said to have planar symmetry.

 The other fields are allowed to depend on both of the
spacelike coordinates. Namely, $A_i=A_i(r,\theta)$, $\Phi=\Phi(r,\theta)$ and
$\Psi=\Psi(r,\theta)$.
As shown explicitly in the Appendix, the field equations admit solutions
where the functional dependence of the fields upon $\theta$ is nontrivial,
 even though the severe restrictions imposed by the 
symmetry of the metric. Such restrictions follow directly from the
Einstein equations (\ref{einsteineq}).  The functional dependence
of the fields on $r$ and $\theta$ has to be such that the total
energy momentum tensor (EMT) results independent on $\theta$. That
is to say, the right hand side of equation (\ref{einsteineq}) must
depend on the radial coordinate only, even though it also involves
functions and derivatives with respect to the angular coordinate.
These compatibility conditions will be automatically satisfied by
any given solution of the field equations.

The task of finding solutions is more easily accomplished by separating
variables through the ansatz
\begin{eqnarray}
{\bf A}(r,\theta)&=& A_1(r)\, A_2(\theta){\bf d}t\, , \label{gaugeansatz}\\
\Phi(r,\theta)& =& \Phi_1(r)\Phi_2(\theta)\, ,\label{phiansatz}\\
 \Psi(r,\theta) &=&\Psi_1(r)\Psi_2(\theta)\, , \label{psiansatz}
\end{eqnarray}
 where the new functions $A_1(r)$, $A_2(\theta)$,
$\Phi_1(r)$, $\Phi_2(\theta)$, etc., depend only on one variable, as
indicated.
This choice for the gauge field ${\bf A}$
excludes 3D spacetimes with magnetic charge such as
the cases considered in Refs. \cite{Hirschwelch,cataldo,dias2},
but it is sufficiently general for the purposes of the present
analysis.

The field equations (\ref{einsteineq})--(\ref{psieom}) then give seven
equations for the unknowns $F$, $A_1$, $A_2$, $\Phi_1$, $\Phi_2$, $\Psi_1$
and  $\Psi_2$. The relevant equations can be cast into
the form presented in the system (\ref{eom1a})--(\ref{eom7a})
in the Appendix.
Some of the equations can be  easily integrated and we are left with the
following set of equations
\begin{eqnarray}
& &  {F^{\prime} \over2}\left({1\over r}+
{\Phi_1^{\prime} \over\Phi_1}\right)-3\alpha^2 +
{A_1 ^\prime}^2+ {F\over r}{\Phi_1^{\prime} \over\Phi_1} +
{\pm k^2\over r^2} + {1\over r^2}
{g^2\over\Phi_1^2}=0\, ,\label{eom1}\\
& & A_1^{\prime}= {q_0 \over\Phi_1\,r}\, , \label{eom3_1}\\
& &   { A_2}={\rm constant}\, , \label{eom3_2}\\
 & & \Phi_1 = p_0 r \, , \label{eom5_1}\\
& &\ddot\Phi_2 =\pm k^{2}\Phi_2 \, , \label{eom5}\\
& & \Psi_1 = {\rm constant}\, ,\label{eom6_1}\\
& &  \dot{\Psi}_2 = g \Phi_2\, ,
\label{eom6}
\end{eqnarray}
where $k^2$, $q_0$, $p_0$ and $g$ are integration constants,
and the prime stands for the derivative with respect to $r$.

Function $\Phi_1(r)=p_0 r $ gives the dependence of the dilaton upon the
coordinate $r$. According to equation (\ref{phiansatz}), the arbitrary constant
$p_0$ can be made equal to unity

As discussed in the Appendix, the constant $\Psi_1(r)$ can also be made
equal to unity without loss of generality, since it is absorbed by
$\Psi_2(\theta)$. Moreover, equations (\ref{eom3_1}) and
(\ref{eom3_2}) tell that
the gauge field ${\bf A}$ is a function of $r$ only. Also, the constant
$A_2$ can be normalized to unity.

The above equations furnish the complete  solutions for
all the unknown fields. $\Phi$ and $\Psi$ are given by (\ref{eom5}) and
(\ref{eom6}), respectively. Equations
(\ref{eom3_1}) and (\ref{eom3_2}) then furnish  $A_1$ which specifies the
electromagnetic potential in its entirety (it is spherically  symmetric).
With these functions at hand we  then finally solve equation (\ref{eom1})
and check the second Einstein equation, (\ref{eom2a}) of the Appendix,
for compatibility.
The result is that the above ansatz  admits five different classes of solutions
characterized by a particular dilaton field and by a particular topology, 
which we enumerate and nominate as follows:

\begin{description}
\item[(i)] {\it Spherically symmetric dilaton  and spherical 3D black holes}:
$\Phi(r,\theta) = c_0\,r$;

\item[(ii)] {\it Non-spherically symmetric dilaton and planar 3D black holes-I}: 
$\Phi(r,\theta)=r (c_0 \theta+ c_1)$;

\item[(iii)] {\it Circular dilaton function - 3D Reissner-Nordstr\"om-AdS black
holes}: $\Phi(r,\theta)=r (c_0 \sin(k\theta) +c_1 \cos (k\theta))$;

\item[(iv)] {\it Circular dilaton function and planar 3D black holes-II}:
$\Phi(r,\theta)=r (c_0 \sin(k\theta) +c_1 \cos (k\theta))$;

\item[(v)]     {\it  Hyperbolic dilaton function and planar black holes-III}:
$\Phi(r,\theta) =r (c_0 \sinh(k\theta) +c_1 \cosh (k\theta))$;
\end{description}
$c_0$ and $c_1$ being arbitrary integration  constants which  assume
different values in each one of the three different cases. 
In the cases (i) and (iii), $\theta$ is a periodic coordinate
($\theta\in [0,2\pi]$), while
in (ii), (iv) and (v), it is nonperiodic ($-\infty< \theta <+\infty$).

The names spherical, non-spherical, circular and hyperbolic are used in 
reference to the functional form of the dilaton field $\Phi(r,\theta)$.
The name
Reissner-Nordstr\"om-AdS is given in reference to those
well known black holes our new solutions can be related to.
Other interesting naming is suggested by the topology of the
corresponding 4D black holes, as explained in the following sections, where
we comment on each one of these cases.

As we have already mentioned, the case $\Phi =$ constant
gives the BTZ black hole solution (see the Appendix).
This black hole has been extensively studied in
the literature and we do not consider it here.

For future reference we write here the contributions to the EMT 
from the fields ${\bf A}$, $\Phi$ and $\Psi$
\begin{eqnarray}
& &\left(T\right)^t_t = {T_r}^r= {{A_1^\prime}^2} -{{\Phi_1}^\prime\over
\Phi_1}\left({F\over r} + {F^\prime\over2}\right) -
{1\over r^2}{\ddot\Phi_2\over\Phi_2} - {1\over r^2} { {\dot\Psi_2}^2\over
{\Phi_2}^2} \,, \\
& &\left(T\right)^r_\theta =\left(T_{A}\right)^\theta_r{r^2 F}=
{1\over r^2}{\dot\Phi_2\over\Phi_2}
\left({{ {\Phi_1}^\prime\over\Phi_1} -{1\over r}}\right)
\, , \\
  & &\left(T\right)^\theta_\theta = - {{A_1}^\prime}^2
-{1\over r^2}{{\dot\Psi_2}^2\over\Phi_2^{2}}
+{{\Phi_1}^\prime\over\Phi_1}F^\prime \, , \label{sphemt}
\end{eqnarray}
the other components of the EMT being zero.
The dependence of the EMT on the angular coordinate is through the
second derivative of the dilaton filed with respect to that variable only.
From equations (\ref{eom5}) and (\ref{eom6}),
it follows that the EMT is a continuous function of $\theta$.
 Moreover, using  the result from
equation
(\ref{eom5_1}) it is seen that $T_r^\theta=0$, and all the other non-zero
components of the energy-momentum tensor are continuous functions of the
$r$ coordinate only, as expected. Also, the Lagrangian dependence on
$\theta$ is carried completely by the dilaton field $\Phi_2(\theta)$.

Now we investigate in some detail the various different solutions which follow
from the above model.

\section{Spherical black hole solutions}
\label{sectspherical}
 This case is  $\Phi =r\,$.
The result for the other scalar field is  $\Psi$ = constant,
 which is obtained from
equation (\ref{sphersol}) in the Appendix by choosing $c_0=c_1=0$. 
It should be noted that considering the angular coordinate 
as a normal azimuthal angle,  the general scalar fields given by
 equations (\ref{sphersol})
are not periodic, and thereby not single-valued functions.
The 3D dilaton field  corresponds to a metric field in the
4D spacetime, which has to be a nonzero well defined function of the 
coordinates.
 Hence, in order to avoid possible problems
due to the multivalued character of the dilaton field $\Phi$ 
we have to  take $c_0=0$.    In fact,  the condition
$\Phi_2(\theta)=\Phi_2(\theta+2\pi)$ implies $c_0=0$. 

The scalar field $\Psi$ corresponds to a gauge potential in 4D spacetime
and, as it is well known from gauge field theory,
 classically it does not have any direct physical meaning.
However, in 3D spacetime, once we have fixed the dilaton field to be
spherical, the only gauge freedom left to the scalar potential $\Psi$
is an arbitrary additive constant. Therefore, by imposing the periodicity 
condition $\Psi(\theta) = \Psi(\theta +2\pi)$ it follows $g=0$. The result
is $\Psi$ = const., which can be made equal to zero.

 The metric and nonzero fields in this case are
\begin{eqnarray}
  ds^2&=& -\left(\alpha^2 r^2-{2m\over r}
-{q^2 \over  r ^2}\right)dt^2 +
{dr^2 \over \alpha^2 r^2-{2m\over r} +{q^2\over r^{2}}} +  r^2
d\theta^2\,, \label{sphermetric}\\
{\mathbf A}& =& -{q\over r}{\mathbf d} t\, ,\label{Aspherical}\\
\Phi&=& r\, . \label{Phitor}
\label{Psispherical}
\end{eqnarray}

Once the coordinates are defined in the intervals $-\infty <t <+\infty$, 
$0\leq r <+\infty$, and $0\leq\theta\leq 2\pi$, 
the present solution is a particular  case of the 3D charged
  black hole studied in  \cite{zkl},
 obtained  through dimensional reduction of the 4D toroidal rotating black
  hole. The properties of this black hole follow  by putting $a=0$ and $g=0$
 in the 3D  black hole studied there. The resulting spacetime has spherical
(circular) symmetry. For the sake of completeness,
and for future comparison, we summarize here the   main properties of such a
spacetime. 

It is straightforward to show
that $m$, is the mass, $q$ is the electric gauge charge, source to the 
field ${\bf A}$.  The other charges and
the angular momentum are zero.

The Ricci and Kretschmann curvature scalars are, respectively,
\begin{eqnarray}
R&=& -6\alpha^2 -{2q^2\over r^4}\, , \label{riccisc}\\
K&=& 12\alpha^4 + {8\alpha^2 q^2\over r^4}+ {24m^2\over r^6}
-{64m q^2\over r^7} + {44q^4\over r^8}\, ,
 \label{riemsq}
\end{eqnarray}
showing  that  there is a singularity at $r=0$.

  The horizons of (\ref{sphermetric}) are given by the real
roots of the equation
\begin{equation}
\alpha^2 r^4 - 2m r +q^2 =0\, . \label{sphhor}
\end{equation}
In the analysis of horizons, the relevant function is $ Descr=   27m^4
-16q^6\alpha^{2}$ and we have three different cases:
 (i) If  $Descr>0$ equation (\ref{sphhor}) has two real positive roots
corresponding
 to  two horizons, the event horizon at $r_+$ and the Cauchy or inner horizon
 at $r_-$. The singularity at $r=0$
 is enclosed by both horizons.
The spacetime can then be extended through the horizons till the singularity.
It represents a 3D static black hole.
(ii) $Descr=0$ --
the solution is the extreme black hole spacetime. There is only one
horizon at $r =r_+=r_- = \sqrt[3]{m/2\alpha^2}\,=\sqrt[4]{\frac43
\frac{q^2}{\alpha^2}\,}$.  The singularity is  hidden (to
external observers) by the horizon.   Geodesic inward lines end at the
singularity $r=0$. (iii) $Descr<0$ --  this solution has no horizons
and represents a naked singularity.

\section{Planar 3D black hole solutions - I}
\label{sectplanar}

There is a second possible interpretation for the solution presented
in equation (\ref{sphersol}) of the Appendix, besides the
spherical black hole of the preceding section.
Namely, the case  where the  spatial sections ($t$ = constant) of the
spacetime have the topology of a plane.

If we assume the ranges of the spatial coordinates as
$0 \leq r < \infty$, and $-\infty < \theta <+\infty$,
then the spacetime is a ``planar" 3D black hole. I. e., 
the topology of $C_\theta$, the spacetime section 
defined by $t=$ constant and $r=$ constant,
is {\bf R} (the real line), which in the 2D spatial section
of the spacetime looks like a one-dimensional wall. 
 
In such a case, it is useful defining a new coordinate $z = L\theta$,
with $L$ being a positive constant carrying dimensions of length
(an interesting choice is to take $L=1/\alpha$ as
 in Ref. \cite{lz1}). 
The scalar fields are then of the form
\begin{eqnarray}
 \Phi&=& r (c_1z+ c_o)\,, \label{phiplanar}\\
\Psi &=& g\left(c_1z^{2} /2 +c_oz\right)+c_2\, ;\label{psiplanar}
\end{eqnarray}
and the relevant metric coefficient is 
\begin{equation}
\displaystyle{F= \alpha^2 r^2-{2m\over r}-{q^2 +g^2\over  r ^2}}\,. 
\label{fplanar}
 \end{equation}
Both of the scalar fields $\Phi$ and $\Psi$, as well as the metric field $F$,
are well behaved functions of the coordinates (except at the singularity
 $r=0$).

As far as the energy momentum tensor is concerned, there are no 
discontinuities like
strings or domain walls in the spacetime.
Moreover, the physical quantities associated to the fields $\bf A$, 
$\Phi$ and $\Psi$, are also well defined everywhere (except at the
singularity). 
The total mass and charges are infinite, but the parameters 
$m$, $q$, etc., are finite (per unit length) quantities.
To see that explicitly let us 
calculate, for instance, the mass of the black hole. The boundary $S_1$ 
[see equation (\ref{massdef})] in the present 
case coincides with the one-dimensional region $C_\theta$ in the limit
$r=\longrightarrow\infty$, and is  not compact.  
Therefore, in order  to avoid
infinite quantities during calculations we select just a finite region of
the boundary between $z=z_{1}$ 
and $z=z_{2}>z_1$, and put $L= z_2-z_1$.
Equation (\ref{massdef}) then gives
$$
 M_L = {L_3\over 4\pi} m\int_{_0}^{\,^L}\!
{c_1 z}\,{dz\over L} = mL\,  ,$$
where we have chosen $L_3= 8\pi/c_1$.  The total mass is infinite since 
$L\longrightarrow\infty$. However the mass per unit length $M_L /L=m$, which
appears in the metric, is a well defined finite quantity. Similarly, the other
charges per unit length of the black hole can be determined (see also the 
next section).

The curvature invariants are the same
as for the spherical black hole considered in the previous section, given
in equations (\ref{riccisc}) and (\ref{riemsq}), with $q^2$ replaced by
 $q^2+g^2$. Namely,
\begin{eqnarray}
R&=& -6\alpha^2 -{2(q^2+g^2)\over r^4}\, , \label{riccisc1}\\
K&=& 12\alpha^4 + {8\alpha^2 (q^2+ g^2)\over r^4}+ {24m^2\over r^6}
-{64m(q^2+g^2)\over r^7} + {44\left(q^2+g^2\right)^2\over r^8}\, .
 \label{riemsq1}
\end{eqnarray}
The singularity at $r=0$ is also a one-dimensional wall. Because
of the singularity, the spacetime cannot be extended to
the region $r< 0$.

The horizons of the planar spacetime are given by the real
roots of the function $F$ given by equation (\ref{fplanar}): $
\alpha^2 r^4 - 2m r +q^2+ g^2 =0\, .$
This solution can then be called a planar black hole (or a black wall) 
spacetime. 
Here, the relevant function is $ Descr=   27m^4
-16(q^2+g^2)^3\alpha^{2}$ and we have three different cases, as in section 
\ref{sectspherical}: (i) Black hole with event and Cauchy horizons 
($Descr>0$), (ii) Extreme black hole ($Descr=0$), (iii) and naked singularity
($Descr<0$).

\section{3D Reissner-Nordstr\"om-AdS black hole solutions}
\label{sectrnads}

The solution in the case
we choose $\Phi=r\left(
c_0\sin k\theta+c_1\cos k\theta\right)$ is
\begin{eqnarray}
\fl ds^2 &=& -\left(k^2+\alpha^2 r^2 -{2m\over r}+{q^2 +g^2\over
  r^2}\right)\, dt^2  + {dr^2\over k^2+ \alpha^2 r^2 - {2m\over r} + {q^2
  +g^2\over r^2}} + r^2 d\theta^2 \, ,\label{3dRNmetric}\\
\fl {\bf A} &=& -{q\over r} {\bf d}t\, ,
\label{3dRNgauge}\\
  \fl \Phi&=& r\left(c_0\sin k\theta+c_1\cos k\theta\right)\, ,
 \label{3dRNdilaton}\\
\fl \Psi &= & {g\over k}\left(-c_0\cos k\theta+{c_1}\sin k\theta
\right)+ c_2\,
\label{3dRNpsi}
\end{eqnarray}

If $\theta$ is interpreted as the azimuthal angle, 
then the topology of $C_\theta$ (the subspace with
$t$ = const., $r$ = const.) is the circle $S^1$, 
and the solution is a 3D  
analogous to the 4D Reissner-Nordstr\"om-AdS  (RNAdS) black hole. 

The spacetime is compact in $\theta$ direction, $\theta$ being an
 azimuthal angle ($\theta$ and $\theta+2n\pi/k$, $n\in Z$, represent
the same set of points in the spacetime). 
Therefore, the scalar fields are continuous 
periodic functions of the coordinate $\theta$. 
The non-negativity of $\Phi^2$ is also guaranteed.
For simplicity, and with no loss o generality, we may choose $c_0=0$, 
$c_1=1$ and $c_2=0$.
Moreover, it is possible to make $k=1$ by a re-parameterization of the 
coordinates: $kt\mapsto t$, $r/k\mapsto r$, $k\theta \mapsto \theta$.
The resulting metric is
exactly the equatorial plane of the 4D Reissner-Nordstr\"om-AdS
 black hole, justifying the name used here.
 Both of the scalar fields are continuous periodic functions of
the azimuthal angle $\theta$.
Hence, the length of a closed
spacelike curve on which $t=$ constant and $r=$ constant is $ 2\pi r$, and
the spacetime may have (circular) horizons depending 
on the relative values of the parameters $m$, $q$, and $g$,
 and has a polynomial singularity at $r=0$ (see below).
  
It is worth mentioning once more that the second derivative of the
dilaton field is a continuous function of $\theta$. This fact is
important for the continuity of the energy momentum tensor
(\ref{sphemt}) which depends explicitly on $\ddot \Phi$ and $\dot
\Psi$. This can also be seen from equations (\ref{eom5}) and
(\ref{eom6}), from what we see that if $\Phi_2$ is a continuous
function of $\theta$, it holds also for $\ddot \Phi_2$ and $\dot
\Psi_2$, implying the continuity of the EMT 
(see e. g. \cite{fiziev}). 

The present 3D RNAdS black hole has several interesting properties
and we study them in some detail in the following. The 
spacetime metric is
 given by (\ref{3dRNmetric}), and
 the physical interpretation of parameters $m$, $q$ and $g$
is given below. The parameter $k^2$ is  kept in the equations, 
even though it can be made equal to unity.

\subsection{Metric and charges:}

In order to investigate the main properties of such a spacetime
 we start computing its global charges.

To define the mass for the metric (\ref{3dRNmetric}) we note that
the one-dimensional boundary ${\cal S}_1$ is defined by the
hypersurface $t={\rm constant}$, $r={\rm constant}$, in the 
asymptotic limit $r\longrightarrow\infty$.
The induced metric on ${\cal S}_1$, as embedded in $D_2$,
$\sigma_{ab}$ is obtained  from (\ref{3dRNmetric}) by putting $dt=0$ and
$dr=0$. Thus, $a,b=2$ and $\sigma_{ab}=\sigma_{22}\equiv\sigma= r^2$.

Comparing equations (\ref{3dRNmetric}) to (\ref{splitmetric}) and using  equation
(\ref{kappadef}) we get the extrinsic curvature of ${\cal S}_1$ modified by
the dilaton $ \kappa^\Phi= -2\sqrt{\Phi^2F/ r^2\,}$,
where $F$ is given  by $F=\alpha^2 r^2+ {k^2} -{2m\over r} +{q^2+g^2\over
  r^2}$. Substituting $\kappa^\Phi$  into (\ref{massdef}) and
integrating over the infinite boundary ${\cal S}_1$ we get the mass of the 3D
black hole
\begin{equation}
 M = {2 L_3\over 4\pi} m \int_{0}^{^{\pi\over k}}\!\!{\cos
 {k\theta}\,\,}\,d\theta = m\,  ,\label{massknads}
 \end{equation}
where we have put $L_3= \pi k$.
By comparison, it is seen that this 3D mass is the  same as  the mass of the
4D Reissner-Nordstr\"om-AdS black hole.

 Since metric (\ref{3dRNmetric}) is static,  it
follows from (\ref{momentumdef}) that  the angular momentum
is zero, $J =0 \,$.

The electric charge comes from  equation (\ref{echarge1}) and is
$ { Q}_e= q\, . $

The full
dilaton field $\Phi$ and
the background dilaton field $\Phi_0$ are equal
$\Phi=\Phi_0= r\,\cos k\theta$,  implying that the dilaton charge is zero.

The charges for the scalar field $\Psi$
are given by equations (\ref{qpsi}) and (\ref{qpsidual}).
In the case under consideration we obtain $
{ Q}_\Psi =0$, and $ {\tilde Q}_\Psi = g\,.$

The above metric has horizons at points where $F(r)=0$ possibly
indicating the presence of a black hole. This is, in fact, the 3D
Reissner-Nordstr\"om-AdS black hole, i.e., a static charged black
hole immersed in an asymptotically AdS spacetime

\subsection{Singularities, horizons, and causal structure:}

The causal structure of the
spacetime given by metric (\ref{3dRNmetric}) is very similar to the
4D Reissner-Nordstr\"om-anti de Sitter black hole. In order to see
this, we show here the singularities and horizons of such a spacetime.

We start computing
the Ricci and the Kretschmann scalars. The result is
the same as in the section \ref{sectspherical}, given in equations
 (\ref{riccisc}) and (\ref{riemsq}).
Thus, there is a singularity at $r=0$ ($R$ and $K$ diverge at $r=0$).
 
The solution
has totally different character depending on whether $r>0$ or $r<0$. The
important black hole solution exists for $r>0$ which case we analyze now.

The horizons are given by the solutions of the equation
\begin{equation}
\Delta = \alpha^2r^4 + k^2r^2 -2 m\,r +q^2+g^2 =0 \, .\label{horrnads}
\end{equation}
By restricting the analysis to $r\geq 0$, the solutions of the above
equation can be classified into three different cases:

\begin{description}
 \item[(i)] If $m^2\left[k^6 +27\alpha^2m^2 -36 k^2\alpha^2(q^2+
   g^2)\right] > ( q^2+g^2)\left[k^4-4\alpha^2 (q^2+g^2)\right]^2$, equation
(\ref{horrnads}) has two real positive roots, $r_+$ and $r_-$, corresponding
respectively to an event horizon and a Cauchy (inner) horizon. The spacetime
is then a 3D spherical black hole whose geodesic and
causal structures are the same as for the equatorial plane of the 4D
Reissner-Nordstr\"om-anti de Sitter black hole. Singularity at $r=0$ is
hidden by the horizon.

\item[(ii)] In the extreme case,
parameters $\alpha^2$, $ m$ and $q$ are related by the constraint $
 m^2\left[k^6 +27\alpha^2 m^2 -36 k^2\alpha^2( q^2+ g^2)\right] = (
 q^2+g^2)\left[k^4- 4 \alpha^2( q^2+g^2)\right]^2\, . $
This is the black hole  in which
the two horizons coincide and are given by
\begin{equation}
r_+=r_-= {k^2 m +12m\,\alpha^2( q^2+g^2)
\over k^2+18\alpha^2 m^2-4k^2\alpha^2 (q^2+g^2) }\, .
\label{extremekn}
\end{equation}

The singularity is hidden by the null horizon at $r=r_+=r_-$, and the causal
structure is the same as the equatorial plane of the 4D extreme
 Reissner-Nordstr\"om-AdS black hole.

\item[(iii)] For $m^2\left[k^6 +27\alpha^2 m^2 -36 k^2\alpha^2
  (q^2+g^2)\right] < ( q^2+ g^2)\left[k^4-4\alpha^2(q^2+ g^2)\right]^2$,
the metric is well behaved over the whole
range from the singularity $r=0$  to the asymptotic limit $r\longrightarrow
\infty$. There are no horizons and the spacetime is a naked singularity.

\end{description}

From the above description,
the Penrose diagrams, with the inherent topology and causal structure
of spacetime,  can easily be drawn.

\subsection{The $\alpha=0$ cases: The 3D Reissner-Nordstr\"om and the
Schwarzschild black holes}

A simple but still interesting 3D spacetime that follows from the
previous solutions is the 3D black hole obtained from (\ref{3dRNmetric})
in the case $\alpha^2=0$,
\begin{eqnarray}
& &ds^2 = -({k^2}-{2m\over r}+ {q^2+g^2\over r^2}) dt^2
+{dr^2\over {k^2}-{2m\over r}+{q^2+g^2\over r^2}} +
r^2 d\theta^2\, ,\label{redRN} \\
\end{eqnarray}
with $\bf{A}$, $\Phi$ and $\Psi$ given respectively by equations
(\ref{3dRNgauge}), (\ref{3dRNdilaton}) and (\ref{3dRNpsi}).
This result can also be seen as the hypersurface $\theta =\pi/2$ of the
4D Reissner-Nordstr\"om  (RN) black hole. Its singularities, horizons
and causal
structure are the same as the equatorial plane of the 4D RN black hole.

For $q^2+g^2\leq m^2$, metric (\ref{redRN}) has two horizons at $r=r_\pm$,
 given by $|k|\, r_\pm=m\pm \sqrt{m^2 -q^2 -g^2\,}$, corresponding
 respectively to event and Cauchy horizons. As in the case $\alpha^2\neq 0$,
there is a singularity at $r=0$  which is
hidden (to external observers at $r\longrightarrow\infty$) by the event
horizon at $r=r_+$.

Whenever the condition $q^2+g^2 >  m^2$ is satisfied, the spacetime
is a naked singularity.

If $q^2+g^2=0$, the solution
obtained from (\ref{redRN}) is a 3D
Schwarzschild spacetime, with the same causal structure,
horizon and singularity as the 4D Schwarzschild black hole.

\subsection{The uncharged case, $g^2 +q ^2 =0$}

If $q^2+g^2=0$, the solution
obtained from (\ref{3dRNmetric}) is a 3D
Schwarzschild-AdS spacetime, with the same causal structure,
horizon and singularity as the 4D Schwarzschild-AdS black hole
(also known as Kottler spacetime)
\begin{eqnarray}
ds^2 &= &-\left({k^2}-{2m\over r} +\alpha^2r^2\right) dt^2 +
{1\over{k^2}-{2m\over r} +\alpha^2r^2 }dr^2
+r^2d\theta^2 \label{schads} \, ,   
\end{eqnarray}
where $\Phi$ is given by (\ref{3dRNdilaton}), and the other fields being zero.

If we also have $m=0$, the solution  is the  (3D) anti-de Sitter metric
with no other charges nor fields besides the dilaton and the cosmological
constant.

\section{Planar 3D  black holes - II}
\label{sectplanar1}

The solution given by equation
(\ref{3dRNmetric})--(\ref{3dRNpsi}) admits a 
different kind of topology besides the spherical
one considered in the preceding section.

In order to see that, define a new 
coordinate $x=L k\theta$ with  $-\infty\leq x<\infty$, $L$ being 
an arbitrary constant carrying dimensions of length,
 and choose scalar fields as in equations (\ref{3dRNdilaton}) and
(\ref{3dRNpsi}):
\begin{eqnarray}
  \Phi&=& r\left(c_0\sin(x/L)+c_1\cos (x/L)\right)\, ,
 \label{3dRNdilaton1}\\
 \Psi &= & {gL}\left(-c_0\cos (x/L)+{c_1}\sin (x/L)
\right)+ c_2\, . \label{3dRNpsi1}
\end{eqnarray}
The metric and gauge fields are given by equations 
(\ref{3dRNmetric}) and (\ref{3dRNgauge}), respectively.

 The resulting spacetime is then
a planar black hole, analogous to the case reported in section
\ref{sectplanar}. 

 The mass and charges can be defined only as
linear densities along the infinite boundary (The line $C_\theta$,
 at the asymptotic region $r\longrightarrow \infty$). 
For instance, the mass within 
a length $L$ of the boundary is 
$M_L=\dspst{{2L_3\over4\pi} m\int_{-\pi L\over2 }^{\pi L\over 2}
\cos{x\over L}\,dx=m L}$ 
(we took $L_3=\pi$). As in  section \ref{sectplanar}, 
the parameter $m$ is then the mass per unit length, while $q$ 
and $g$ are the charge densities associated respectively to the 
gauge field $\bf A$ and to the scalar field $\Psi$. 

The local geometric properties and the causal structure of this solution is,
however, very similar to the 3D RNAdS black hole studied in
section \ref{sectrnads}, and we do not discuss the details here.

\section{Planar 3D black hole solutions - III }
\label{secthyper}

The general form of the solution obtained when we choose 
$\displaystyle{\ddot\Phi_2\over \Phi_2}= +k^2 >0$
is [see equations (\ref{hypersol}) in the Appendix]
\begin{eqnarray}
\fl ds^2 &=& \left(-k^2+\alpha^2r^2 -{2m\over r}  + {q^2 +g^2\over
  r^2}\right) dt^2 + {dr^2\over \left(-k^2+\alpha^2r^2 -{2m\over r} +
  {q^2 +g^2\over  r^2}\right)}+r^2d\theta^2\, ,\label{hypermetric}\\
\fl  {\mathbf A} & =& -{q\over r}\,{\mathbf d}t\, ,\label{Ahyper}\\
\fl \Phi&=& r\, \left(c_0\sinh k\theta+c_1\cosh k\theta\right)\,
 \label{Phihyper}\\
\fl \Psi &= & {g\over k}\left(c_0\cosh k\theta+c_1 \sinh k\theta
\right) +c_2\, ,\label{Psihyper}
\end{eqnarray}
 where $m$, $c_0$, $c_1$, $c_2$, $g$, $q$ and $k$ are generic
integration constants. The coordinates  can be normalized such
that $k^2=1$. The solution corresponds to the hypersurface
$\rho =$ constant of the topological black hole with genus $\tt{g}  >1$
of reference \cite{vanzo} (see also
 \cite{klemm_mor_vanzo}).  The space surface
 $t=$ constant, $\rho=$ constant of such a 4D black hole has the topology of
a hyperbolic two-space $H^2$.
The functional form of $\Phi$ also justifies the adjective hyperbolic used
in the present case.

The spacetime cannot be compactified in $\theta$ 
direction, because the dilaton field would be a multivalued
function.

The choice $c_0=0$, $c_1>0$, and $\theta$ in the range
 $-\infty \leq \theta < +\infty$, assures the dilaton field is a 
continuous and  non-zero function of the
coordinates. 
 Moreover,
 $\ddot \Phi$ and $\dot \Psi$ are continuous functions of
$\theta$, implying the continuity of the EMT, what guarantees
the basic properties of a physical spacetime
(see e. g. \cite{fiziev}).

As in the case of section \ref{sectplanar}, mass, and 
charges are infinite, but the linear densities of these quantities
are finite and well defined.
 
The locus $r=0$ is a true spacetime singularity, since the Ricci
 and Kretschmann scalars are the same as in section \ref{sectplanar}
(given respectively by equations (\ref{riccisc1}) and (\ref{riemsq1})).
Moreover, there are metric singularities (horizons) at points where
\begin{equation}
\Delta = \alpha^2r^4 - r^2 -2m\,r +q^2+g^2 =0 \,  .\label{horhyper}
\end{equation}
Depending on the relative values of the parameters, $\alpha^2$, $m$,
$q$ and $g$, there may be two horizons and the solution is
an asymptotically anti-de Sitter black hole.
 Even though these very interesting properties,
 a more exhaustive analysis of such a black hole will not 
be presented in
this paper.

\section{Rotating charged 3D  black holes}
\label{sectrotating}

\subsection{The rotating metric}

The circular black holes discussed in sections
\ref{sectspherical} and \ref{sectrnads} can be put to rotate.
In order to add angular momentum to the spacetime we perform
 a local coordinate transformation which mixes
time and angular coordinates as follows  (see e.g.
\cite{lemos1,lz1,Sa_Lemos,HorWel})
\begin{eqnarray}
 t &\mapsto& t-\frac{\omega}{\alpha^2} \theta \:,
                                       \nonumber  \\
 \theta &\mapsto& \gamma \theta \:,
                                       \label{transf-rot}
\end{eqnarray}
where $\gamma$ and $\omega$ are constant parameters.

The next step is substituting the transformation (\ref{transf-rot}) into
 the equations of each one of the above static solutions. For instance,
taking (\ref{3dRNmetric})--(\ref{3dRNpsi}) we find
\begin{eqnarray}
d s^{2}& =& -F\left( dt - {\omega\over\alpha^2} d\theta\right)^2 +
\gamma^2 r^2 d\theta^2  + \frac{dr^2}{F}  \, , \label{rotmetric1}\\
 {\bf  A} & =& -{q\over r} \left( {\rm\bf d}t
-{\omega\over\alpha^2}{\rm\bf d}\theta\right) \, ,
\label{rotgauge} \\
\Phi  &= &r \cos\left(\gamma\,k\theta\right)\, ,\label{rotdilaton}\\
\Psi &= & -{g\over k} \sin\left(\gamma\,k\theta\right)\, , \label{rotpsi}
\end{eqnarray}
where 
\begin{equation}
F= k^2 +\alpha^2r^2 -{2m\over r} +{q^2+g^2\over r^2}\, , \label{F1}
\end{equation}
and we chose $c_0=0$, $c_1=1$ and $c_2=0$.

Even though we used the solution (\ref{3dRNmetric})--(\ref{3dRNpsi})
explicitly, the constant $k^2$ is kept explicitly in the metric, so that the
two particular cases $k^2=0$ (section \ref{sectspherical}), 
  and $k^2> 0$ (section \ref{sectrnads}) are both contained in
the analysis. The case $k=0$ is obtained by taking the appropriated 
limit in the above equations. In that case, one also has to put $g=0$, i.e., 
$\Psi=0$, and $\Phi=r$ for all $\theta\in (0,2\pi)$, as shown in 
section \ref{sectspherical}.

Planar black holes such as those reported in sections \ref{sectplanar},
\ref{sectplanar1}
and \ref{secthyper} cannot be put to rotate.

 Let us
mention that parameters $\gamma$ and $\omega$ are not independent
and one of them can be redefined by a re-parameterization of the
coordinates. An interesting choice for $k^2\neq 0$ is
$\gamma^{2}(1-k^2)=\frac{\omega^2}{\alpha^2}$, what ensures the
asymptotic form of the metric for large $r$ is such that the
length of a closed curve $t=$ constant, $r=$ constant, is exactly
$2\pi r$. Another possible and simple choice is to take
$\gamma=1$, leaving  $\omega$ free. This last choice is used
throughout this paper because it applies also to the case $k^2=0$.

 Analyzing the Einstein-Rosen bridge of the
static solution one concludes that the spacetime is not simply
connected which implies that the first Betti number of the manifold is
one, i.e., closed curves encircling the horizon cannot be shrunk
to a point.
Therefore, transformations (\ref{transf-rot}) generate a new metric
because they are not permitted global coordinate transformations
 \cite{Stachel}.
Metrics (\ref{3dRNmetric}) and
(\ref{rotmetric1}) are distinct, for they can be locally mapped into
each other but not globally.

\subsection{The global charges}
  In the spacetime associated to the
metric (\ref{rotmetric1}) we choose a region $\cal{M}$ of
spacetime bounded by $r={\rm {\rm constant}}$, and two space-like
surfaces $t=t_{1}$ and $t=t_{2}$. The region of the spacetime $t={\rm
constant}$, $r={\rm constant}$, is the one-dimensional boundary
${\cal S}_1$ of the two-space $D_2$.
The boundary of $\cal{M}$,
$\partial \cal M$, in the present case consists of the product of
${\cal S}_1$ with timelike lines ($r={\rm constant }\,
,\theta={\rm constant}$) joining the surfaces $t=t_1$ and $t=t_2$,
and these two surfaces themselves.
 The metric $\sigma_{ab}$ is obtained from
 (\ref{rotmetric1})
by making $dt=0$ and $dr=0$, while the
two-space $D_2$ metric $h_{ij}$ ($i,j=1,2$) is obtained by putting $dt=0$.
The timelike and spacelike unit vectors $t^i$ and $n^i$ are, respectively,
$t^i = {\delta_0^i\over N}-{N^\theta\delta_2^i\over N}$, $n^i =
\sqrt{F}\delta_1^i$. Metric (\ref{rotmetric1}) admits the two Killing vectors,
a spacelike $\eta_\theta$ and a timelike $\eta_t$, needed in order to define
mass and angular momentum.

Comparing metric (\ref{rotmetric1}) to (\ref{splitmetric}) and
using (\ref{kappadef})  we get the following expression for the
extrinsic curvature of ${\cal S}_1$,
\begin{equation}
{\kappa^\Phi}= -{1\over 2}\sqrt{F\Phi^2} \,\,
\left[{2r \left(1- {\omega^2\over\alpha^2}\right)-{\omega^2\over\alpha^4}
 \left({2m\over   r^2}-2{q^2 +g^2\over r^3}\right) \over r^2
\left(1-{\omega^2\over\alpha^2}\right)- {\omega^2\over\alpha^4}\left(k^2
-{2m\over r} +{q^2 +g^2\over r^2}\right)} +{2\over r}\right]\, ,
 \label{rotkappa}
\end{equation}
where $F$ and $\Phi$ are given respectively
by (\ref{F1}) and (\ref{rotdilaton}).
Let us mention that the above equation holds for  both of the
classes of spacetimes presented in sections \ref{sectspherical} and
 \ref{sectrnads}.

To build $\delta\left(\kappa^\Phi\right)=\kappa^\Phi-
\left(\kappa^\Phi\right)_{\!o}$ we take
$\kappa^\Phi$ from equation (\ref{rotkappa}),
  which was obtained from the full solution given in equations
(\ref{rotmetric1})--(\ref{rotpsi}).
The extrinsic curvature of the background spacetime,
$\left(\kappa^\Phi\right)_{\!o}$,
 is obtained from the 3D spacetime with no black hole
present, and follows
from the same relation (\ref{rotkappa}) by choosing $m=0$, $q=0$ and $g=0$.
Then,  substituting $\delta\left(\kappa^\Phi\right)$
into (\ref{massdef}), and taking the limit $r\longrightarrow\infty$,
the mass of the toroidal 3D black hole is finally obtained,
\begin{equation}
 M=m\left(1+ {3\over2\alpha^2}{\omega^2\over 1- {\omega^2\over\alpha^2}}
 \right)\,,\label{rotmass}
\end{equation}
 where $m$ is the mass of the static black hole, and to simplify
we have put $L_3 = \pi|k|$ for $k^2>0$,  and  $L_3 =2$ in the case $k^2=0$.
Assuming  $0\leq{\omega^2\over\alpha^2} < 1$, we see that the rotating
black hole  mass is larger than the mass of the original static black hole.
 The additional mass, $\delta M={3m\over2\alpha^2}{\omega^2\over1-
{\omega^2\over\alpha^2}}$, depends explicitly on the  rotation parameter
$\omega$ and can be viewed as being generated by the motion of the 3D system.

To calculate the angular momentum we compare metric (\ref{rotmetric1})
to (\ref{splitmetric}) and use equation (\ref{jaydef}) which gives
\begin{equation}
j^{\Phi}_i\eta_\theta^i= {\omega\over \alpha^2}\sqrt{\Phi^2\,}
{\left(F' r-2F\right) \over \sqrt{r^2 -{\omega^2F\over\alpha^2}}}\, ,
\label{jay1} \end{equation}
where $F'$ indicates the derivative with respect to $r$. From the above
result and
equation (\ref{momentumdef}) the angular momentum follows
\begin{equation}
J = {3\over 2} m {\omega\over\alpha^2}\, .\label{amgularjay}
\end{equation}
 However, it should be emphasized that the angular momentum as defined by
 equation (\ref{jaydef}) is shown to be a conserved quantity only in the case
 $k=0$, when
the dilaton field is rotationally invariant \cite{creighton-mann}.  The
dilaton field must be constant on orbits of the Killing vector
$\eta_\theta=\frac{\partial}{\partial\theta}$, which excludes the case
corresponding to $k^2\neq 0$.

Other nonzero charges are the same as for the static black hole,
$Q_e\ = q,$ and $\tilde Q_\Psi = g.$

\subsection{Causal Structure of the Charged Rotating  Spacetime}

We can  see that  the metric for the charged rotating spherically
symmetric 3D asymptotically anti-de Sitter spacetime, given in 
(\ref{rotmetric1}), has the same singularities and horizons as the
metric of the non-rotating spacetime given by (\ref{3dRNmetric}).
Of course, there are differences and we will explore them here.

The coordinate transformation (\ref{transf-rot}) does not change
local properties of the spacetime. Hence,  the Kretschmann scalar $K$ is the
same as for the non-rotating spacetime, and so are the singularities.

The horizons of the rotating metric (\ref{rotmetric1}) are located
at points where $F(r)=0$. Hence, the horizons are still given by
the solutions of equation (\ref{horrnads}), the same as for the
static spacetime.

The infinite redshift surfaces, which are given by the zeros of  the metric
coefficient $g_{tt}$, are also the same for both the rotating and the static
spacetime. Such a surface corresponds to each one of the roots of the equation
$F(r)=0$, and coincides with the horizons.

The causal structure of static and rotating spacetimes are very similar
indeed.
The differences appear when one studies the existence of closed
timelike curves (CTCs).

To study closed timelike curves (CTCs) we first note that the angular Killing
vector ${\partial}_{\theta}$ has norm given by
$\displaystyle{{\partial}_{\theta}.{\partial}_{\theta}
= g_{\theta\theta} = r^2-{\omega^2}F/\alpha^4.}$
There are CTCs for $g_{\theta\theta}<0$. The
radii for which $g_{\theta\theta}=0$ are given by the solutions of  the
equation
\begin{equation}
\alpha^2r^4\left(1-{\omega^2\over\alpha^2}\right) -{\omega^2\over
  \alpha^2}k^2\, r^2 + 2{\omega^2\over\alpha^2}m\, r-{\omega^2\over\alpha^2}
  \left(q^2+ g^2\right) =0 \,  .\label{rctc}
\end{equation}
It is  a quartic equation and one can easily find its zeros in terms of the
other parameters. To avoid writing a new set of equations for these zeros, we
just comment on some interesting cases one may find.

To begin with, we consider firstly the case $k^2>0$ and choose the other
parameters in such a way that the spacetime has two horizons, $r_-$ and $r_+$.
 Since we fixed $0\leq{\omega^2\over\alpha^2}<1$, the solutions of last
 equation are threefold depending on the discriminant
\begin{equation*}
\fl D= m^2\left\{{\omega^2\over  \alpha^2}k^4\left(3-
 {\omega^2\over\alpha^2}
k^6\right)+18\alpha^2\left(1-{\omega^2\over\alpha^2}\right)
\left[2 \left(q^2+g^2\right)\left(1+{\omega^2\over\alpha^2}k^6\right)
-3m^2 {\omega^2\over\alpha^2}k^4\right]\right\} \end{equation*}
\begin{equation}\label{descr}
-2\left(q^2+g^2\right)\left[1 + 4\alpha^2k^2\left(q^2+g^2\right)
  \left(1-{\omega^2\over\alpha^2}   \right) \right]^2\, .
\end{equation}
The three possibilities for the real roots of (\ref{rctc}) are:
 (i) If $D > 0$ there are three positive roots; (ii) for $D=0$ two
 positive roots; and (iii) one positive root if $D>0$. We assume
$\alpha^2 >0$ and $m >0$.

Firstly, if $D<0$, equation (\ref{rctc})  has just one real positive zero, $r_0$,
 say. This happens in particular when
$\omega^2/\alpha^2$  is small
compared to unity, and for a large range of the other parameters. In this
case, $r_0$ is smaller than the horizon $r_+$ for all the range of the other
parameters. The CTCs are located in the region $r < r_0$, so
there are CTCs just inside the event horizon. Moreover, there are CTCs
outside the inner horizon, $r_0 > r_-$, just when the charge is
very small compared to the mass, otherwise the CTCs happen just inside
the inner horizon.

In the case $D=0$, equation (\ref{rctc}) has two  positive
roots. Let us name them $r_0$ and $r_1$ and assume $r_1> r_0$.
Similarly to the case $D<0$ above, there are CTCs
in the region $r < r_0$. However, in the region which the radial coordinate
$r$ assumes values between $r_0$ and $r_1$, $r_0< r< r_1$, there are two
possibilities depending on the values of the discriminant
$$
D_1 = 6k^2\alpha^2\left(q^2+g^2\right)\left(1-{\omega^2\over \alpha^2}\right)
- \left[{\omega^2\over 3\alpha^2} k^4 + 4\alpha^2 \left(1-{\omega^2\over
  \alpha^2}\right) m\right]^2 .$$
For $D_1> 0$ the CTCs are restricted to the region $r <r_0$ and $r_0$ is
inside the horizon $r_+$.  If $D_1<0$
 there are CTCs in all the region $0\geq r< r_1$, and $r_1$ may be larger than
 the horizon.

The third case, $D> 0$ is when
equation (\ref{rctc}) has three positive solutions, $r_0$, $r_1$ and
$r_2$. There are CTCs in both of the regions $r_1< r < r_2$ and $r <
r_0$. In the region $r_0<r<r_1$ there are no CTCs.
 The values of $r_1$ and $r_2$ depend strongly  on the parameter $\omega^2$,
 and for $\omega^2/\alpha^2$ sufficiently close to unity they can be both
 larger than the event horizon.

 Now, for $k^2= 0$, equation (\ref{rctc}) has just
one positive root $r_0$, say, which depends strongly on the values of the
charges $q$ and $g$, and vanishes when $g^2+ q^2=0$.
Moreover, the CTCs are
restricted to the region inside the event horizon.
For $r> r_0$ there are no CTCs.

\section{Conclusions}
\label{sectconclusions}

Using  a 3D dilaton-gravity Lagrangian, with additional scalar and
gauge fields, inspired in the dimensional reduction of the 4D
Einstein-Maxwell Lagrangian
 we obtained new 3D static black hole solutions.

Assuming an ansatz where the spacetime is static and  has
circular symmetry, but allowing the other fields
to be anisotropic, we show that the field equations do not
impose the circular symmetry on the fields. In particular, the
dilaton field needs not be isotropic.  By using spherical
coordinates $(r,\theta)$, the anisotropy of the static dilaton
field shows up through its explicit dependence on the angular
coordinate. There are five different forms of the dilaton as a
function of $r$ and $\theta$, which are compatible with the
symmetry of the spacetime. Namely, a constant; a function of $r$
only; a function of $r$ times a linear function of theta;
a function of $r$ times a circular function of $\theta$;
and a function of $r$ times a hyperbolic function of $\theta$.

 As a consequence, we
find at least six different families of black holes, each one 
characterized by a different dilaton field.

The first is the well known BTZ family of black holes, which follows
when we choose the dilaton to be constant.

Secondly, when the dilaton is isotropic (depends only on the radial
coordinate), it is found a 3D spherical black hole solution 
which corresponds to the 4D cylindrical (or toroidal) black hole
studied in  \cite{zkl} with one
dimension suppressed.

The dilaton can also be a linear function on of $\theta$, which 
is not a compact coordinate. The 
solution may be interpreted as a planar asymptotically
AdS black hole. 

Another very interesting
 black hole solution  is found when we choose the dilaton dependence
upon $\theta$ to be the cosine function. This implies in a 3D 
spherically symmetric spacetime which
corresponds to the equatorial plane of the 4D Reissner-Nordstr\"om-AdS black
hole.  
Changing the topology of the spacetime in the $\theta$ direction to be the
real line, the fifth class of solutions is found as a variant to the 3D RNAdS
case.

The last class of 3D black hole spacetimes, which has planar topology,
 is obtained in the case the dilaton is
a hyperbolic function of the angular coordinate $\theta$. Such a spacetime
corresponds to the 4D charged hyperbolic black hole of references 
\cite{vanzo,klemm_mor_vanzo,peldan}
with one dimension suppressed.

The spherical black holes can be put to rotate by a coordinate transformation
which mixes time and angle, but only in cases when the dilaton is isotropic the
angular momentum is a conserved quantity.

\section*{Acknowledgments}
One of us (V.T.Z.) thanks Centro Multidisciplinar de Astrof\'\i sica at
Instituto Superior T\'ecnico (CENTRA-IST) for a grant and for hospitality
while part of this work has been done. A.S.M. thanks the Brazilian
Agency CAPES for a grant.
We are indebted to J. P. S. Lemos for many interesting discussions and 
encouragement.

\section*{Appendix: General equations and solutions}

The 3D metric is initially assumed to be static but without rotational
symmetry, which in Schwarzschild-like coordinates reads

\begin{equation}
ds^2= -f^2\, dt^2 +g^2\, dr^2 +h^2\, d\theta^2\, ,\label{ametric}
\end{equation}
where $f$, $g$, and $h$ are functions of the coordinates $r$ and
$\theta$ only.
The gauge field ${\bf A} $ is chosen in the form
${\bf A}= A_t(r,\theta) {\bf d}t$.

The Lagrangian ${\cal L}$ defined from equation (\ref{lagr3d1})
 is then
\begin{eqnarray}
\fl {\cal L} &=& {L_3\over 16\pi}\Phi f gh\left[
 {2\over h^2} \left({\dot f\over f}{\dot h\over h}-
{\dot f\over f}
{\dot g\over g} + {\dot g \over g} {\dot h \over h}-{\ddot f\over f}
 - {\ddot g\over g} \right)
+ {2\over g^2}\left({f^\prime \over f}
{g^\prime \over g}+{g^\prime \over g} {h^\prime \over h}
- {\dot f\over f}{\dot h\over h}- {f^{\prime\prime} \over f}
-{h^{\prime\prime} \over h} \right) \right.\nonumber \\
\fl & &\left. +6\alpha^2
+{2\over f^2}
\left( {{A_t^\prime}^2\over g^2}+ {{\dot{A}_t}^2\over h^2}\right)
-{2\over \Phi^2}\left( {{\Psi^\prime}^2\over g^2}+
{{\dot \Psi}^2\over h^2}\right)\right]\, ,\label{lagrap}
\end{eqnarray}
where a prime  denotes partial derivative with
respect to $r$, and a  dot  denotes partial derivative with respect
to $\theta$.

Now we calculate the equations of motion (EOM) as defined in section
\ref{lagrangian}.

{From} the metric (\ref{ametric}) we get the following non-zero components of
the Einstein tensor
\begin{eqnarray}
G^t_t & =& {1\over  h^2} {\ddot{g} \over g}+ {1\over g^2}
{h^{\prime\prime} \over h} -{1\over g^2} { g^\prime \over g }
{ h^\prime\over h}-{1\over h^2}{\dot{g}\over g}{\dot{h} \over h}\, ,\\
G^r_r & &= {1\over  h^2} {\ddot f \over f}
-{\dot f\over f }{\dot h \over h^3}
+{1\over g^2}{f^\prime\over f}{h^\prime \over h}\, , \\
G^r_\theta & =& G^\theta_r {h^2\over g^2}=-{1\over  g^2}
{\dot{f}^\prime \over f}
+{1\over g^2} {\dot g\over g}{f^\prime\over f }
+{1\over g^2}{\dot f\over f}{g^\prime\over g}\, ,\\
G^\theta_\theta & =& {1\over  g^2} {f^{\prime\prime} \over f}
-{1\over g^2} { f^\prime\over f }
{ g^\prime\over g}+ {1\over h^2}{\dot f\over f}{\dot g\over g}\, ,
\end{eqnarray}

Before writing explicitly the EMT components and the EOM for the fields,
we restrict the metric to have spherical symmetry we have $f=1/g=F(r)$,
$h=r$. In such a case, the nonzero components of the Einstein tensor
are $G_t^t=G_r^r= F^\prime/2r$, $G_\theta^\theta= F^{\prime\prime}/2$.

Moreover, to simplify the analysis, the other fields are split into
the product of two independent functions of each one of the coordinates $r$
and $\theta$, $A_t=A_1(r)\, A_2(\theta)$, $\Phi = \Phi_1(r)\,
\Phi_2(\theta)$, $\Psi= \Psi_1(r)\, \Psi_2(\theta)$. Substituting this
ansatz into the Einstein  equations (\ref{einsteineq}) we get
\begin{eqnarray}
 & &\fl G_t^t ={F^\prime\over 2r} =3\alpha^2 - { {{A_1}^\prime}^2 {A_2}^2}
- {\dot{A_2}^2 {A_1}^2\over r^2 F} - F {{\Phi_1}^{\prime\prime}\over
  \Phi_1} -  {F^\prime\over 2}{{\Phi_1}^\prime\over \Phi_1}
 -{F\over r}{{\Phi_1}^\prime\over\Phi_1}
-{1\over r^2}{\ddot{\Phi_2}\over \Phi_2}  \, \nonumber \\
& &
- F {{{\Psi_1}^\prime}^2 \Psi_2^2\over{\Phi_1}^2{\Phi_2}^2}
-{1\over r^2}{{\Psi_1}^2{\dot\Psi_2}^2\over{\Phi_1}^2{\Phi_2}^2} \, ,
\label{eintt}\\
& &\fl G^r_r = {F^\prime\over 2r} = 3\alpha^2 - { {{A_1}^\prime}^2 {A_2}^2}
+ {\dot{A_2}^2 {A_1}^2\over r^2 F} -{F^\prime\over 2}
{\Phi_1^\prime\over \Phi_1} -{F\over r}{{\Phi_1}^\prime\over\Phi_1}
-{1\over r^2}{\ddot{\Phi_2}\over \Phi_2}  \, \nonumber \\ &&
+ F {{{\Psi_1}^\prime}^2 {\Psi_2}^2\over{\Phi_1}^2{\Phi_2}^2}
-{1\over r^2}{{\Psi_1}^2{\dot\Psi_2}^2\over{\Phi_1}^2{\Phi_2}^2} \, ,
\label{einrr}\\
& &\fl G^r_\theta = 0 = -2A_1A_2 A_1^\prime\, \dot A_2
+{F}{\Phi_1^\prime\over \Phi_1}{\dot\Phi_2\over \Phi_2}
-{F\over r}{\dot\Phi_2\over \Phi_2} + 2F{\Psi_1\Psi_2\over \Phi_1^2\Phi_2^2}
{\Psi_1^\prime} \dot\Psi_2\, ,\label{einrth}\\
 & &\fl G^\theta_\theta  =  {F^{\prime\prime} \over 2} = 3\alpha^2 -
 {{{A_1}^\prime}^2 {A_2}^2} +{\dot{A_2}^2 {A_1}^2\over r^2 F} - F
 {{\Phi_1}^{\prime\prime}\over\Phi_1} -  {F^\prime}
 {{\Phi_1}^\prime\over \Phi_1} -  F {{{\Psi_1}^\prime}^2
   \Psi_2^2\over{\Phi_1}^2{\Phi_2}^2}
+{1\over r^2}{{\Psi_1}^2{\dot\Psi_2}^2\over{\Phi_1}^2{\Phi_2}^2} \, ,
\label{einthth}
\end{eqnarray}
where now the prime and the dot denote total derivative with respect to $r$ and
 $\theta$, respectively.

Variation of the action $S=\int{{\cal L}\, d^3x}$
with respect to $\Phi$ gives the constraint
\begin{equation}
 R+6\alpha^2 -{\mathbf F}^2 +{2}\Phi^{-2} ({\bf \nabla} \Psi)^2 =0
 \, ,\label{phieom0}
\end{equation}
which in the present case assumes the  form
\begin{equation}
\fl {2F^\prime \over r}+{F^{\prime\prime}} -6\alpha^2
-{A_1^\prime}^2{A_2}^2 - {{A_1}^2{\dot{A}_2}^2\over r^2 F}
-{2\over {\Phi_1}^2{\Phi_2}^2}\left( {\Psi_1^\prime}^2{\Psi_2}^2
+{\Psi_1}^2{\dot \Psi_2}^2\right) = 0\, . \nonumber
\end{equation}
The above constraint is in fact one of the equations of motion.
 Variation of the action
with respect to $g_{ij}$, $A_t$ and $\Psi$ completes the set of EOM, given
respectively by equations (\ref{einsteineq}), (\ref{gaugefieldeom}) and
(\ref{psieom}). Equation (\ref{phieom}) is obtained  by substituting
(\ref{einsteineq}) into equation (\ref{phieom0}).
The system of equations of motion as written in section \ref{lagrangian} does
not admit solutions without dilaton. In such a case,
equation (\ref{phieom0}) is not part of the system of EOM, and
neither is equation (\ref{phieom}), so that the case $\Phi=$ constant
has to be considered separately (see below).

The sole Maxwell equation for the gauge field ${\bf A}$, equation 
(\ref{gaugefieldeom}), is
\begin{equation}
A_2\left({A_1^{\prime\prime}} + A_1^\prime{1\over r} +
A_1^\prime{\Phi_1^\prime\over \Phi_1}\right) +{A_1\over r^2F}
\left({\ddot A_2} +\dot A_2{\dot\Phi_2\over \Phi_2}\right) = 0\, .
\label{max1}
\end{equation}

The equations for the scalar fields $\Phi$ and $\Psi$, (\ref{phieom}) and
(\ref{psieom}), respectively, reduce to
\begin{eqnarray}
 & & -3\alpha^2+ F {{\Phi_1}^{\prime\prime}\over
  \Phi_1} + {{\Phi_1}^\prime\over \Phi_1}\left({F^\prime\over 2}
 +{F\over r}\right)
-{1\over r^2}{\ddot{\Phi_2}\over \Phi_2}
- F {{{\Psi_1}^\prime}^2 \Psi_2^2\over{\Phi_1}^2{\Phi_2}^2}
-{1\over r^2}{\dot{\Psi_1}^2{\Psi_2}^2\over{\Phi_1}^2{\Phi_2}^2}\, \nonumber\\
& & - { {{A_1}^\prime}^2 {A_2}^2}
- {\dot{A_2}^2 {A_1}^2\over r^2 F}=0 \hfill  \, ,\label{phieq1} \\
 & &  {{\Psi_1}^{\prime\prime}\over
  \Psi_1} + {F^\prime\over F}{{\Psi_1}^\prime\over \Psi_1}
 +{1\over r}{{\Psi_1}^\prime\over\Psi_1} -{\Phi_1^\prime\over
  \Phi_1}{\Psi_1^\prime\over \Psi_1}
+{1\over F r^2}\left({\ddot{\Psi_2}\over \Psi_2}
-{\dot{\Psi_2}\over{\Psi_2}}{\dot{\Phi_2}\over{\Phi_2}}\right)=0\, .
\label{psieq1}
 \end{eqnarray}

Equations (\ref{eintt})--(\ref{psieq1}) are the basic equations to be solved.
We shall see, however, that this system of equations can be significantly
simplified after a careful analysis.

{F}rom  equations (\ref{phieq1}) and (\ref{eintt})
 we find
\begin{equation}
F^\prime\left({1\over r} -{\Phi_1^{\prime}\over \Phi_1}\right) =0 \, .
\label{eom0a}
\end{equation}
The solution of this equation is twofold: (i) $F(r)=$ constant and
$\Phi_1(r)$ arbitrary, or (ii) $F(r)$ arbitrary and $\Phi_1(r)= c_0 r$,
where $c_0$ is a nonzero constant.
The case (i) is not interesting, because it leads to a flat spacetime.
 Hence, we take the second possibility $\Phi_1(r) =c_0 r$ and
leave $F(r)$ to be determined by the  set of remaining equations.

Equations (\ref{eintt}) and
(\ref{einrr}) yield
\begin{equation}
 F{\Phi_1^{\prime\prime}\over\Phi_1}+A_1^2{{\dot A_2 }^2\over r^2 F^2}
 + \Psi_2^2 {{\Psi_1^\prime}^2\over\Phi_1^2\Phi_2^2} =0 \, ,\label{eom0b}
\end{equation}
which, together with equation (\ref{eom0a}), implies $A_2=$ constant and
$\Psi_1=$ constant. These constants
can be made equal to unity, since they are absorbed by the functions
 $ A_1$ and $\Psi_2$, respectively, in the definitions $A_t(r,\theta)=
A_1(r)A_2(\theta)$ and $\Psi(r,\theta)= \Psi_1(r)\Psi_2(\theta)$. 
Hence, the gauge potential $A_t$ depends only on the coordinate $r$, while 
$\Psi$ depends only on the coordinate $\theta$.

By rewriting the field equations including these
results, the system of seven field equation
(\ref{eintt})--(\ref{psieq1}) may then be replaced by the following
equivalent set
\begin{eqnarray}
& &  {F^{\prime} \over2}\left({1\over r}+
{\Phi_1^{\prime} \over\Phi_1}\right)-3\alpha^2 +
{A_1 ^\prime}^2+ {F\over r}{\Phi_1^{\prime} \over\Phi_1} +
{1\over r^2}{{\ddot\Phi}_2\over\Phi_2} + {{\Psi_1}^2\over r^2}
{\left.{\dot\Psi}_2\right.\!^2\over{\Phi_1}^2{\Phi_2}^2}=0\, ,\label{eom1a}\\
 & & {1\over2}F^{\prime\prime}+ {F^{\prime} \over2}\left({1\over
  r}+{3\Phi_1^{\prime}\over\Phi_1}\right)- 6\alpha^2  -
 {F\over r}{\Phi_1^{\prime} \over\Phi_1}
+ {1\over r^2}{{\ddot\Phi}_2\over\Phi_2}=0  \, ,\label{eom2a} \\
& & A_1^{\prime}\, \frac{\Phi_1^\prime}{\Phi_1} +A_1^{\prime\prime}+
\frac{1}{r}A_1^{\prime} =0\, ,           \label{eom3a}\\
& & {\dot A_2}=0\, , \label{eom4a}\\
& & \Phi_1^{\prime\prime} = 0\, ,\label{eom5a}\\
 & & {\dot{\Phi}_2\over\Phi_2}\left({\Phi_1'\over\Phi_1}-{1\over
r}\right) =0  , \label{eom5b} \\
& & \Psi_1^\prime =0\, ,\label{eom6a}\\
& & \ddot{\Psi}_2-\dot{\Psi}_2{\dot{\Phi}_2\over\Phi_2}=0\, .\label{eom7a}
\end{eqnarray}

Even though there are eight equations for seven unknowns, there is
no inconsistency, since the two first equations of the last system
are not independent (see below).

The first two equations follow from the Einstein equations
(\ref{eintt}) and (\ref{einthth})
after using the results of (\ref{eom0a}) and (\ref{eom0b}).
Equation (\ref{eom3a}) follows by substituting
$\dot A_2=0$ into Maxwell equation (\ref{max1}).
Equation (\ref{eom5b}) is obtained from Einstein equation
(\ref{einrth}) after using the results implied by (\ref{eom0b}).

We observe that  equations (\ref{eom5a}) and (\ref{eom5b})
 are equivalent to the EOM
for the dilaton, and includes the case $\Phi=$ constant as
a particular solution. Notice also that (\ref{eom2a}), (\ref{eom4a}),
(\ref{eom5a}), (\ref{eom5b}) and  (\ref{eom6a}) imply the Einstein equation
(\ref{einrr}). The last two equations are equivalent to  the EOM  for $\Psi$
(see equation (\ref{psieq1})).

Function $\Phi_1(r)$ gives the dependence of the dilaton upon the
coordinate $r$, which according to equations
(\ref{eom5a}) and (\ref{eom5b}) is twofold:
(i)  $\Phi_1(r)=$ constant, in which case  $\Phi_2$ must also
 be a constant; (ii)
$\Phi_1(r)=r$, in this case  $\Phi_2(\theta)$ is not fixed by
equation (\ref{eom5b}).

Now, equation (\ref{eom2a}) requires that
\begin{equation}
{\ddot\Phi_2\over \Phi_2} =\pm k^{2} \, , \label{phi2eq}
\end{equation}
which together with (\ref{eom1a}) gives
\begin{equation}
{\dot\Psi_2\over \Phi_2} =g \, , \label{psi2eq}
\end{equation}
$k$ and $g$ being arbitrary constants.
The last equation is consistent with (\ref{eom7a}) and, in fact, replaces it.

The general solution for $\Phi$ in the case of
spherical symmetry we are considering here can be split in four
different particular cases, corresponding to the four different
possibilities that follow from equations (\ref{eom5a}) and
(\ref{phi2eq}). Namely,
(i) $\Phi_1^\prime=0$ and $\ddot\Phi_2 = 0 $, (ii) $ r \Phi_1^\prime=\Phi_1$
 and $\ddot\Phi_2 =0$,
(iii) $ r\Phi_1^\prime=\Phi_1$ and $\ddot\Phi_2 = +k^2 >0$, and (iv)
  $ r \Phi_1^\prime=\Phi_1$ and $\ddot\Phi_2 = -k^2 <0$.

Taking the results for the fields ${\bf A}$, $\Phi$ and $\Psi$
 into account we are left with the following
equations for $F$
\begin{eqnarray}
& &  {F^{\prime} \over2}\left({1\over r}+
{\Phi_1^{\prime} \over\Phi_1}\right)-3\alpha^2 +
{q^2\over r^2\Phi_1^2}+ {F\over r}{\Phi_1^{\prime} \over\Phi_1} +
{k^2\over r^2} + {g^2\over r^2 {\Phi_1}^2}=0\, ,\label{eom1b}\\
 &&{1\over2}F^{\prime\prime} +
{F^{\prime} \over2}\left({1\over
  r}+{3\Phi_1^{\prime}\over\Phi_1}\right)
- 6\alpha^2  - {F\over r}{\Phi_1^{\prime} \over\Phi_1}
+ {k^2\over r^2}=0  \, ,\label{eom2b}
\end{eqnarray}

It is worth noticing that if $\Phi_1^\prime =0$ then
equation (\ref{eom5a}) implies $\dot\Phi_2=0$ which requires
$k^2=0$.

It is straightforward to show that the two above equations are
functionally dependent. For, multiply (\ref{eom1b}) by $r^2$,
differentiate
with respect to $r$ and use the fact that ${r\Phi_1^\prime\over
  \Phi_1}=$ constant to find
$$
{F^{\prime\prime}\over 2}\left( 1+{r\Phi_1^\prime\over\Phi_1}\right)
+{F^\prime\over 2}\left( {1\over r}+{3\Phi_1^\prime\over\Phi_1}\right)-
6\alpha^2\left( {1}+{r\Phi_1^\prime\over\Phi_1}\right)
+2{r\Phi_1^\prime\over \Phi_1}\left({k^2\over r^2}
+{F^\prime\over 2}{\Phi_1^\prime\over \Phi_1}+ {F\over r}
{\Phi_1^\prime\over \Phi_1}\right) =0\, .
$$
This last equation is identical to (\ref{eom2b}) in the two possible
cases, ${r\Phi_1^\prime\over \Phi_1}=1$ and
${r\Phi_1^\prime\over \Phi_1}=0$. The last condition requires $k=0$
in equation (\ref{phi2eq}).

Now we write down the explicit solutions for each field
considering the four different cases for dilaton field
function mentioned above.

First consider $\Phi=$ constant.
 The final solution in this case is
\begin{eqnarray}
F &=& 3\alpha^2 r^2 + 2\left(q^2+ g^2\right)\ln r -M\, ,\nonumber\\
{\mathbf A } &=& q \ln r\ {\bf d} t\, , \nonumber\\
\Phi &=& c_0\, , \nonumber\\
\Psi &=& g \left(c_0\theta+ c_1\right)\, . \label{btzsol}
\end{eqnarray}

The second class of solutions is that related to the choice
$\Phi= r(c_0\theta +c_1)$.  The relevant functions
are
\begin{eqnarray}
F &=& \alpha^2 r^2-{2M \over r}+{q^2+ g^2\over r^2}\, ,\nonumber\\
{\bf A } &=& {2q \over r}\ {\bf d} t\, , \nonumber\\
\Phi&=& r(c_0\theta + c_1)\, , \nonumber\\
\Psi &=& g \left({c_0\over 2}\theta^2 +c_1\theta\right)+c_2\, .
\label{sphersol}
\end{eqnarray}

The third type of solutions are obtained in the case $\ddot\Phi_2 <0$
and is given by
\begin{eqnarray}
F &=& k^2+\alpha^2 r^2-{2m \over r}+{q^2+ g^2\over r^2}\, ,\nonumber\\
{\bf A } &=& {2q \over r}\ {\bf d} t\, , \nonumber\\
\Phi&=& r\left(c_0\sin\, k\theta + c_1\cos\, k \theta\right)\, , \nonumber\\
\Psi &=&-g \left({c_0\over k}\cos\, k\theta -{c_1\over
  k}\sin\, k\theta\right)+c_2\, . \label{rnadssol}
\end{eqnarray}

Finally, the fourth class of solutions follow in the case  $\ddot\Phi_2 >0$
and is of the form
\begin{eqnarray}
F &=& -k^2+\alpha^2 r^2-{2M \over r}+{q^2+ g^2\over r^2}\, ,\nonumber\\
{\bf A } &=& {2q \over r}\ {\bf d} t\, , \nonumber\\
\Phi&=& r\left(c_0\sinh\, k\theta + c_0\cosh\, k \theta\right)\, , \nonumber\\
\Psi &=& g \left({c_0\over k}\cosh\, k\theta +{c_1\over
  k}\sinh\, k\theta\right)+c_2\, . \label{hypersol}
\end{eqnarray}

The solution given  by equations (\ref{btzsol}) is a simple generalization
of the BTZ black hole. The second black hole solution, given in
(\ref{sphersol}), was
found in reference \cite{zkl}. The two last cases, (\ref{rnadssol}) and
 (\ref{hypersol}), are new solutions which may represent 3D black holes.
In particular the case (\ref{rnadssol}) is similar to the 4D
Reissner-Nordstr\"om-AdS black hole.

\section*{References}
\label{biblio}

\end{document}